\documentclass[12pt]{article}
\input xy
\xyoption{all}
\usepackage{graphicx}

\def\sqr#1#2{{\vcenter{\vbox{\hrule height.#2pt
            \hbox{\vrule width.#2pt height#1pt \kern#1pt
                  \vrule width.#2pt}\hrule height.#2pt}}}}
                                                                                
\def\square
 {\mathop{\mathchoice{\sqr{12}{15}}{\sqr{9}{12}}
{\sqr{6.3}{9}}{\sqr{4.5}{9}}}}

\begin{document}

\hfill hep-th/0608056

\vspace{1.5in}

\begin{center}

{\large\bf Derived categories and stacks in physics}

\vspace{0.2in}

Eric Sharpe\\
Departments of Physics, Mathematics\\
University of Utah\\
Salt Lake City, UT  84112\\
{\tt ersharpe@math.utah.edu} \\

$\,$

\end{center}

This is a summary of a talk given at the Vienna homological mirror
symmetry conference in June 2006.  We review how both derived categories
and stacks enter physics.  The physical realization of each has
many formal similarities.  For example, in both cases, equivalences are realized
via renormalization group flow:  in the case of derived categories,
(boundary) renormalization group flow realizes
the mathematical procedure of localization on quasi-isomorphisms,
and in the case of stacks, worldsheet renormalization group flow
realizes presentation-independence.
For both, we outline current technical issues and applications.

\begin{flushleft}
August 2006
\end{flushleft}

\newpage

\tableofcontents

\newpage

\section{Introduction}

For many years, much mathematics relevant to physics (Gromov-Witten
theory, Donaldson theory, quantum cohomology, {\it etc})
has appeared physically in correlation function computations in
supersymmetric field theories.
Typically one can see
all aspects of the mathematics encoded somewhere in the physics,
if one takes the time to work through the details.
In this fashion we have been able to understand the relation of
these parts of mathematics to physics very concretely.

However, more recently we have begun to see a more complicated dictionary,
in which mathematical ideas of homotopy and categorical equivalences map to
the physical notion of the renormalization group.
The renormalization group is a very powerful idea in physics,
but unlike the correlation function calculations alluded to in the
last paragraph, it is not currently technically feasible to follow
the renormalization group explicitly and concretely a finite distance
along its flow.
Unlike what has happened in the past,
we can no longer see all details of the
mathematics explicitly and directly in the physics, 
and instead have to appeal to
indirect arguments to make the connection.

In this note we will outline two such recent examples of pieces of
mathematics in which important components map to the renormalization group
in physics.  Specifically, we will briefly discuss how derived categories and
stacks enter physics.
For information on the mathematics of derived categories,
see for example \cite{weibel,hred,thomasdc}, and
for a more extensive description of how derived categories enter physics,
see the review article \cite{dclec}.
For information on the mathematics of stacks see
\cite{vistoli,gomezstx}.

\section{The renormalization group}

For readers not\footnote{The talk was given to an audience of both
mathematicians and physicists, and we have attempted to make these
notes accessible to both audiences as well.} 
familiar with the notion, the renormalization group is
a {\it semi}group operation on an abstract space of physical theories.
Given one quantum field theory, the semigroup operation constructs new
quantum field theories which are descriptions valid at longer and longer
distance scales.

In particular, under the renormalization group two distinct
theories can sometimes become the same (the semigroup operation
is not invertible).
We have schematically illustrated such a process in the two pictures below.
Although the two patterns look very different, at long distances
the checkerboard on the right becomes a better and better approximation
to the square on the left, until the two are indistinguishable.
\begin{center}
\begin{tabular}{cc}
\includegraphics[width=1.5in,height=1.5in]{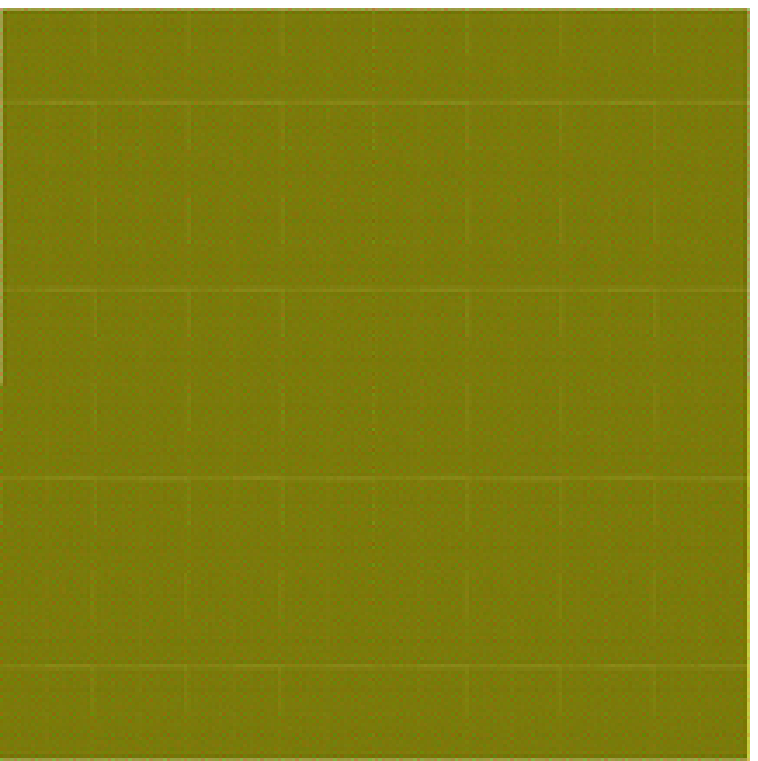} &
\includegraphics[width=1.5in,height=1.5in]{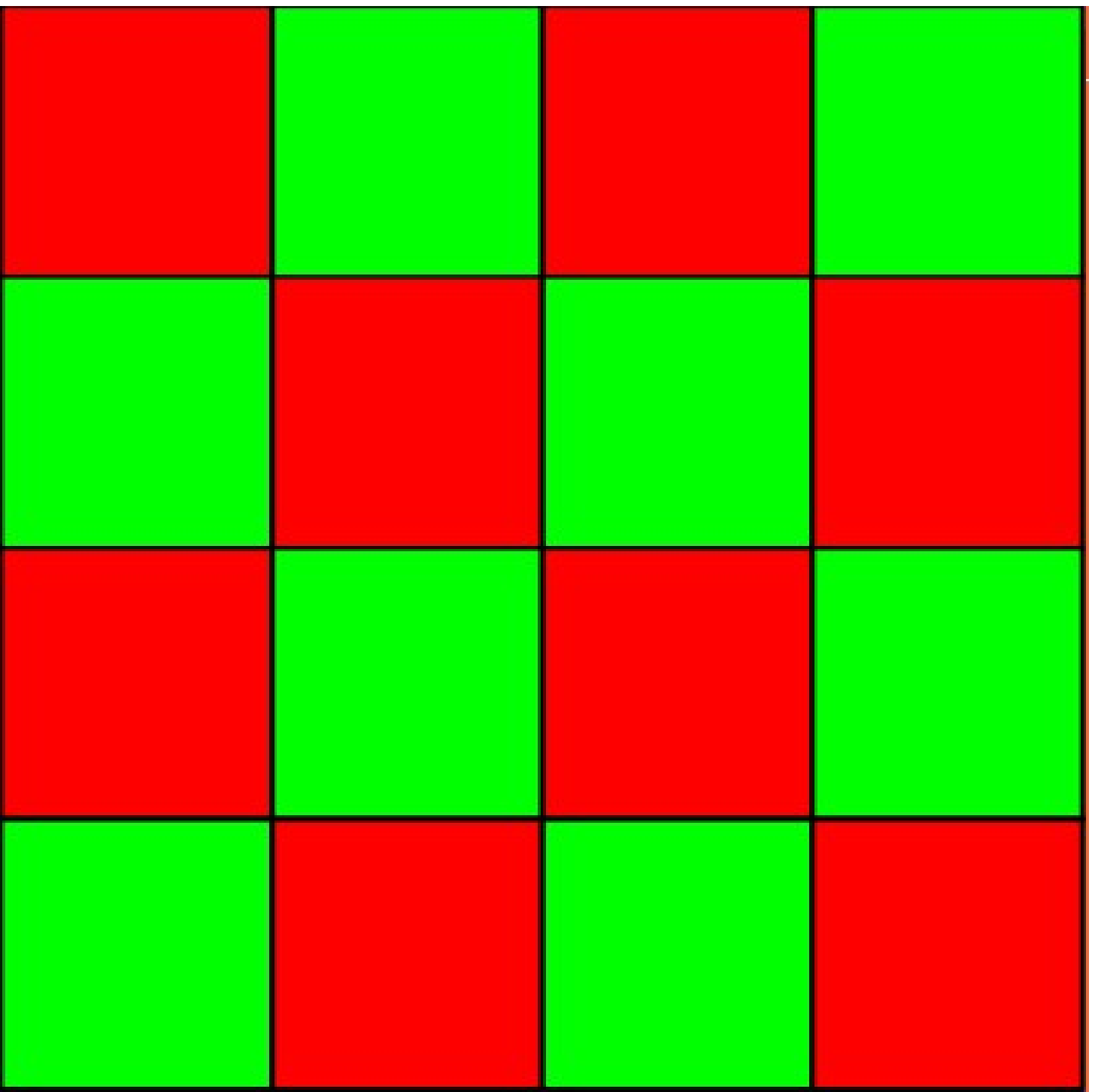}
\end{tabular}
\end{center}

Two theories that flow to the same theory under the renormalization group
are said to be in the same ``universality class'' of renormalization group 
flow.

The renormalization group is a powerful tool but unfortunately we cannot
follow it completely explicitly in general.  The best we can typically
do is construct an asymptotic series expansion to the tangent vector
of the flow at any given point.
Thus, ordinarily we cannot really prove in any sense that two theories
will flow under the renormalization group to the same point.

On the one hand, a mathematical theory that makes predictions for
how different theories will flow, as happens in the application
of both derived categories and stacks to physics,
is making strong statements about physics.
On the other hand, we can only check such statements indirectly, by
performing many consistency tests in numerous examples.

\section{Derived categories in physics}

\subsection{History}

Derived categories entered physics gradually through a succession of
developments.  Before describing the modern understanding, let us take
a moment to review the history.

One of the original motivations was work of Kontsevich,
his ``homological mirror symmetry'' approach to mirror symmetry
\cite{kontsevich}.  Ordinary mirror symmetry is a relation between
two Calabi-Yau's, but Kontsevich's Homological mirror symmetry relates derived
categories of coherent sheaves on one Calabi-Yau to Fukaya categories on
the other.  At the time it was proposed, no physicist had any
idea how or even if derived categories entered physics -- 
his proposal predated even D-branes --  a testament to
Kontsevich's insight.

Shortly after Kontsevich's work, in 1995 Polchinski reminded everyone
of his work on D-branes, and explained their relevance to
string duality \cite{polchinski}.  Although neither derived categories
nor sheaves appeared in \cite{polchinski}, they would later play a role. 

About a year later, Harvey and Moore speculated in \cite{hm} that
coherent sheaves might be a good mathematical description for
some D-branes.  Given that impetus, other authors soon discovered
experimentally that, indeed, mathematical properties of coherent
sheaves at least often computed physical quantities of
corresponding D-branes.  For example, it was discovered empirically
that massless states between D-branes were counted by Ext groups
between the corresponding sheaves, though the complete physical
understanding of why that was the case was not worked out until
\cite{ks,kps,cks}.  

Having understood how sheaves could at least often be relevant to physics
was an important step, but derived categories of sheaves are more
complicated than just sheaves -- derived categories of sheaves
involve complexes of sheaves, not just sheaves.
The next intellectual step was Sen's introduction of antibranes
\cite{sen}, and Witten's realization that Sen's work amounted
to a physical realization of K theory \cite{edktheory} in 1998.

With that insight, the first proposals for how derived categories
enter physics became possible.  Shortly after Witten's introduction
of K theory, it was proposed in \cite{medercat} that the same notion
of antibranes, when applied to sheaf models, could be used to give
a physical realization of derived categories.
Specifically, a complex of branes representing any given object
in a derived category should correspond physically to a set of
branes and antibranes, with the maps in the complex defining a set of
tachyons.  Two objects in the derived category related by
quasi-isomorphism should correspond physically to two sets of
branes and antibranes that are in the same universality class of
renormalization group flow.  In this fashion one could finally begin to have
a physical understanding of Kontsevich's homological mirror symmetry.

Although several papers were written and talks given,
this physical understanding of derived categories
languished in obscurity until Douglas popularized
the same notion two years later in \cite{doug1}.
Douglas also introduced the notion of pi-stability,
which has seen some interest in the mathematics community
(see for example \cite{bridgeland}).

\subsection{D-branes and sheaves}

To lowest order, a D-brane is a pair, consisting of a submanifold
of spacetime together with a vector bundle on that submanifold.
In fact, we will specialize to D-branes in the open string topological
B model, in which case the submanifolds are complex submanifolds and
the vector bundles are holomorphic.
Such data we can describe by the sheaf $i_* {\cal E}$,
where $i$ is the inclusion map of the submanifold into the spacetime,
and ${\cal E}$ is a vector bundle on the submanifold.

In what sense are sheaves a good model for D-branes?
After all, physically D-branes are specified by a set of boundary conditions
on open strings together with Chan-Paton data, which is not the same
thing as a sheaf.
However, we can compute physical quantities (such as massless spectra)
using mathematical operations on sheaves (such as Ext groups),
and it is for this reason that we consider sheaves to be a good model
for D-branes.

In particular, mathematical deformations of a sheaf match
physical deformations of the corresponding D-brane.  This is ordinarily
one of the first tests that one performs given some new mathematical
model of part of physics.

Let us next briefly outline how these sorts of computations are performed.

Massless states in the topological B model are BRST-closed combinations
of the fields $\phi^i$, $\phi^{\overline{\imath}}$, $\eta^{\overline{\imath}}$,
$\theta_i$, modulo BRST-exact combinations.
The fields $\phi^i$, $\phi^{\overline{\imath}}$ are local coordinates on
the target space, and $\eta^{\overline{\imath}}$, $\theta_i$ are
Grassman-valued.  In simple cases, the BRST operator acts as follows:
\begin{eqnarray*}
Q_{BRST} \cdot \phi^i \: = \: 0, & \: &
Q_{BRST} \cdot \phi^{\overline{\imath}} \: \neq \: 0 \\
Q_{BRST} \cdot \eta^{\overline{\imath}} \: = \: 0, & \: &
Q_{BRST} \cdot \theta_i \: = \: 0
\end{eqnarray*}
States are then of the form
\begin{displaymath}
b(\phi)^{\alpha \beta\: \: \: j_1 \cdots j_m}_{
\overline{\imath}_1 \cdots \overline{\imath}_n} \eta^{\overline{\imath}_1}
\cdots \eta^{\overline{\imath}_n} \theta_{j_1} \cdots \theta_{j_m}
\end{displaymath}
where $\alpha$, $\beta$ are ``Chan-Paton'' indices, coupling to two
vector bundles ${\cal E}$, ${\cal F}$.
We can understand these states mathematically by applying the
dictionary (for an open string both of whose ends lie on a submanifold
$S \subseteq X$)
\begin{displaymath}
Q_{BRST} \: \sim \: \overline{\partial}, \: \:
\eta^{\overline{\imath}} \: \sim \: d \overline{z}^{\overline{\imath}}
\: \sim \: TS, \: \:
\theta_i \: \sim \: {\cal N}_{S/X}
\end{displaymath}
Then the states above can be identified with elements of the
sheaf cohomology group
\begin{displaymath}
H^n(S, {\cal E}^{\vee} \otimes {\cal F} \otimes \Lambda^m {\cal N}_{S/X} )
\end{displaymath}

The analysis above is a bit quick, and only applies in special cases.
In general, one must take into account complications such as:
\begin{enumerate}
\item The Freed-Witten anomaly \cite{freeded}, which says that to the
sheaf $i_* {\cal E}$ one associates a D-brane on $S$ with `bundle'
${\cal E} \otimes \sqrt{K_S}$ instead of ${\cal E}$ \cite{ks}.
\item The open string analogue of the Calabi-Yau condition,
which for two D-branes with trivial bundles wrapped on submanifolds
$S$, $T$, becomes the constraint \cite{ks}
\begin{displaymath}
\Lambda^{top} {\cal N}_{S \cap T/S} \otimes
\Lambda^{top} {\cal N}_{S \cap T/T} \: \cong \: {\cal O}
\end{displaymath}
\item When the Chan-Paton bundles have nonzero curvature, the boundary
conditions on the fields are modified \cite{abooetal}; for example,
for line bundles, the constraint can be written
\begin{displaymath}
\theta_i \: = \: F_{i \overline{\jmath}} \, \eta^{\overline{\jmath}}
\end{displaymath}
\end{enumerate}
Taking into account these complications will in general physically realize
a spectral sequence \cite{ks}; 
for example, for D-branes wrapped on the same submanifold
$S$,
\begin{displaymath}
H^n\left(S, {\cal E}^{\vee} \otimes {\cal F} \otimes
\Lambda^m {\cal N}_{S/X} \right) \: \Longrightarrow \:
\mbox{Ext}^{n+m}_X\left( i_* {\cal E}, i_* {\cal F}\right)
\end{displaymath}

Not all sheaves are of the form $i_* {\cal E}$ for some vector
bundle ${\cal E}$ on $S$.  How can one handle more general cases?

A partial answer was proposed in \cite{gs} (before the publication
of \cite{doug1}) and later worked out in more detail in
\cite{dks}, using mathematics appearing in \cite{del}.
The proposal made there was that some more general sheaves can
be understood as mathematical models of D-branes with 
nonzero ``Higgs fields.''  For each direction perpendicular to the
D-brane worldvolume, there is a Higgs field, which one can interpret
as a holomorphic section of ${\cal E}^{\vee} \otimes {\cal E}
\otimes {\cal N}_{S/X}$.  The idea is that we can interpret such 
a section as defining a deformation of the ring action on the
module for ${\cal E}$, yielding a more general module,
meaning a more general sheaf.

A trivial example of this is as follows.
Start with a skyscraper sheaf at the origin of the complex line,
corresponding to a single D-brane at the origin.
The module corresponding to that sheaf is ${\bf C}[x]/(x)$.
Now, consider the Higgs  vev $a$ 
(${\cal E}^{\vee} \otimes {\cal E} \otimes {\cal N}_{S/X} \cong
{\cal O}$ here).  Describe the original module as a generator
$\alpha$ subject to the relation $x \cdot \alpha = 0$,
then define the new module by $x \cdot \alpha = a \alpha$.
That relation is the same as $(x-a) \cdot \alpha = 0$,
and so the new module is ${\bf C}[x]/(x-a)$, which describes
a D-brane shifted from the origin to point $a$.
A Higgs field in such a simple case should translate the D-brane,
so this result is exactly what one would expect.

Similarly, the sheaf ${\bf C}[x]/(x^2)$ (the structure sheaf of
a nonreduced scheme) corresponds to a pair of D-branes over the
origin of the complex line with Higgs field
\begin{displaymath}
\left[ \begin{array}{cc}
0 & 1 \\
0 & 0
\end{array} \right]
\end{displaymath}

We believe this mathematics has physical meaning because massless states
in theories with nonzero Higgs fields can be shown to match Ext groups
between the sheaves obtained by the process above.
Physically, a nonzero Higgs field deforms the BRST operator to the form
\begin{displaymath}
Q_{BRST} \: = \: \overline{\partial} \: + \:
\Phi_1^i \theta_i \: - \: \Phi_2^i \theta_i
\end{displaymath}
where $\Phi_1^i$, $\Phi_2^i$ are Higgs fields on either side of the
open string.
A necessary condition for the topological field theory to be sensible is
that the BRST operator square to zero, which imposes the constraints
that the Higgs fields be holomorphic, and that the different Higgs fields
commute with one another (which ordinarily would be an $F$ term condition
in the target-space theory).
When one computes massless states using the deformed BRST operator above,
one gets Ext groups between the sheaves dictated by the dictionary above.
See \cite{dks} for more information.

\subsection{Derived categories}

There is more to derived categories than just, sheaves.
For example, where does the structure of complexes come from?
Not to mention, where does the renormalization group enter?

First, in addition to D-branes, there are also anti-D-branes.
An anti-D-brane is specified by the same data as a D-brane,
but dynamically a D-brane and an anti-D-brane will try to annihilate
one another.

Furthermore, in addition to antibranes, there are also ``tachyons'' between
branes and antibranes, represented by maps between the sheaves representing
the branes and antibranes.

The dictionary between derived categories and physics can now be stated.
Given a complex
\begin{displaymath}
\cdots \: \longrightarrow \: {\cal E}_0 \: \longrightarrow \:
{\cal E}_1 \: \longrightarrow \: {\cal E}_2 \: \longrightarrow \:
\cdots
\end{displaymath}
we map it to a brane/antibrane system in which the
${\cal E}_i$ for $i$ odd, say, define branes, the other sheaves define
antibranes, and the maps are tachyons \cite{medercat,doug1}.

The first problem with this dictionary is that we do not know how to
associate branes to every possible sheaf:  we can map branes to sheaves
but not necessarily the reverse.

The solution to this problem is as follows.
So long as we are on a smooth complex manifold, 
every equivalence class of objects has a representative in terms
of a complex of locally-free sheaves, {\it i.e.} a complex of bundles,
and we do know how to associate branes to those.

So, for any given equivalence class of objects, we pick a 
physically-realizable representative complex (at least one exists),
and map it to branes/antibranes/tachyons.

The next problem is that such representatives are not unique,
and different representatives lead to different physics.
For example, the sheaf $0$, describing no branes or antibranes,
 is equivalent in a derived category to
the complex
\begin{displaymath}
0 \: \longrightarrow \: {\cal E} \: \stackrel{=}{\longrightarrow} \:
{\cal E} \: \longrightarrow \: 0
\end{displaymath}
which is described by an unstable set of equivalent branes and antibranes.
However, although these two systems are physically distinct,
we believe that after a long time they will evolve to the same
configuration -- the branes and antibranes will completely annihilate.
Such time evolution corresponds to worldsheet boundary renormalization
group flow.

Thus, the proposal is that any two brane/antibrane systems representing
quasi-isomorphic complexes flow to the same physical theory under the
renormalization group.  In other words, the mathematics of derived
categories is providing a classification of universality classes
of open strings.

A proposal of this form can not be checked explicitly -- it is not technically
possible to explicitly follow renormalization group flow.
Thus, we must perform numerous indirect tests, to accumulate
evidence to determine whether the proposal is correct.

One test we can perform is to calculate massless spectra in the
nonconformal theory describing brane/antibrane/tachyon systems and check
that, again, one gets Ext groups.  Let us work through those details.

On the worldsheet, to describe tachyons, we add a term to the boundary,
which has the effect of modifying the BRST operator, which becomes
\begin{displaymath}
Q_{BRST} \: = \: \overline{\partial} \: + \: \sum_i \phi_i^{\alpha \beta}
\end{displaymath}
schematically.
A necessary condition for the topological field theory to
remain well-defined is that $Q_{BRST}^2 = 0$,
which implies that \cite{paulalb}
\begin{enumerate}
\item $\overline{\partial} \phi^{\alpha \beta} = 0$, {\it i.e.} the maps
are holomorphic
\item $\phi_i^{\alpha \beta} \phi_{i+1}^{\beta \gamma} = 0$,
{\it i.e.} the composition of successive maps vanishes, the condition for
a complex.
\end{enumerate}

Furthermore, if $f_{\cdot}: C_{\cdot} \rightarrow D_{\cdot}$ is
a chain  homotopy between complexes, {\it i.e.} if 
\begin{displaymath}
f \: = \: \phi_D s \: - \: s \phi_C
\end{displaymath}
for $s_n:  C_n \rightarrow D_{n-1}$, then $f = Q_{BRST} s$,
and so is BRST exact.  So, modding out BRST exact states will have
the effect of modding out chain homotopies.

As an example, let us compute $\mbox{Ext}^n_{{\bf C}}({\cal O}_D, {\cal O})$
in this language, for $D$ a divisor on the complex line ${\bf C}$.
\begin{displaymath}
0 \: \longrightarrow \: {\cal O}(-D) \: \stackrel{\phi}{\longrightarrow} \:
{\cal O} \: \longrightarrow \: {\cal O}_D \: \longrightarrow \: 0
\end{displaymath}
Relevant boundary states are of the form
\begin{displaymath}
\begin{array}{c}
b^{\alpha \beta}_{0 \overline{\imath}_1 \cdots \overline{\imath}_n}
\eta^{\overline{\imath}_1} \cdots \eta^{\overline{\imath}_n}
\: \sim \: H^n\left( {\cal O}(-D)^{\vee} \otimes {\cal O} \right) \\
b^{\alpha \beta}_{1 \overline{\imath}_1 \cdots \overline{\imath}_n}
\eta^{\overline{\imath}_1} \cdots \eta^{\overline{\imath}_n}
\: \sim \: H^n\left( {\cal O}^{\vee} \otimes {\cal O} \right)
\end{array}
\end{displaymath}
In this language, degree one states are of the form $b_0 + 
b_{1 \overline{\imath}} \eta^{\overline{\imath}}$.
The BRST closure conditions are
\begin{displaymath}
\begin{array}{c}
\overline{\partial} b_0 \: = \: - \phi( b_{1 \overline{\imath}} d
\overline{z}^{\overline{\imath}} \\
\overline{\partial} \left( b_{1 \overline{\imath}} d
\overline{z}^{\overline{\imath}} \right) \: = \: 0
\end{array}
\end{displaymath}
and the state is BRST exact if
\begin{displaymath}
\begin{array}{c}
b_0 \: = \: \phi a \\
b_{1 \overline{\imath}} d \overline{z}^{\overline{\imath}} \: = \:
\overline{\partial} a
\end{array}
\end{displaymath}
for some $a$.
These conditions imply that
\begin{displaymath}
b_0 \mbox{ mod }\mbox{Im } \phi \: \in \:
H^0\left( D, {\cal O}(-D)^{\vee}|_D \otimes {\cal O}|_D \right) 
\: = \:
\mbox{Ext}^1( {\cal O}_D, {\cal O})
\end{displaymath}

Conversely, given an element of 
\begin{displaymath}
\mbox{Ext}^1 ( {\cal O}_D, {\cal O}) \: = \: H^0\left( D,
{\cal O}(-D)^{\vee}|_D \otimes {\cal O}_D\right)
\end{displaymath}
we can define $b_0$ and $b_1$ using the long exact sequence
\begin{displaymath}
\cdots \: \longrightarrow \: H^0\left( {\cal O} \right) \: \longrightarrow \:
H^0\left( {\cal O}(D) \right) \: \longrightarrow \:
H^0\left( D, {\cal O}(D)|_D\right) \: \stackrel{\delta}{\longrightarrow} \:
H^1({\cal O}) \: \longrightarrow \: \cdots
\end{displaymath}
from which we see $b_1$ is the image under $\delta$ and
$b_0$ is the lift to an element of $C^{\infty}({\cal O}(D))$.

More generally, it can be shown that Ext groups can be obtained in this fashion.

Thus, massless spectra can be counted in the nonconformal theory, and they
match massless spectra of the corresponding conformal theory:
both are counted by Ext groups.
This gives us a nice test of presentation-independence of renormalization
group flow, of the claim that localization on quasi-isomorphisms is realized
by the renormalization group.

\subsection{Grading}

Let us next take a few minutes to describe how the grading appears
physically in terms of $U(1)_R$ charges.

For branes and antibranes wrapped on the entire space, the analysis is
straightforward.
The tachyon $T$ is a degree zero operator.
The term we add to the boundary to describe a tachyon is
the descendant
\begin{displaymath}
\int_{\partial \Sigma} [G, T]
\end{displaymath}
where $G$ is the topologically-twisted boundary supercharge.
The operator $G$ has $U(1)_R$ charge $-1$, so $[G,T]$ has charge $-1$.
Now, a necessary condition to preserve supersymmetry is that boundary
terms must be neutral under $U(1)_R$ (otherwise the $U(1)_R$ is broken,
which breaks the ${\cal N}=2$ boundary supersymmetry).
Thus, the Noether charge associated to the $U(1)_R$ symmetry must have
boundary conditions on either side of the boundary-condition-changing 
operator above such that the grading shifts by one.

For lower-dimensional sheaves, on the other hand, the relationship
between the $U(1)_R$ charge and the grading is more subtle.
In particular, the state corresponding to an element of
$\mbox{Ext}^n({\cal S}, {\cal T})$ for two sheaves ${\cal S}$, ${\cal T}$
need not have $U(1)_R$ charge equal to $n$ -- the degree of the Ext group
will not match the charge of the state.
If we build the states as combinations of fields acting on a vacuum,
then the $U(1)_R$ charge of the field combinations will be the 
same as the degree of the Ext group, but the vacuum will make an
additional contribution to the $U(1)_R$ charge which will spoil the
relationship.  In particular, if the two sheaves do not correspond
to mutually supersymmetric branes, then the charge of the vacuum need
not even be integral.  This mismatch is very unlike closed strings,
where typically the vacuum charge contribution precisely insures
that the total $U(1)_R$ charge {\it does} match the degree of corresponding
cohomology.  This mismatch was known at the time of \cite{medercat},
and has been verified more thoroughly since (see {\it e.g.}
\cite{dclec}), though it is often misstated in the literature.

\subsection{Generalized complexes}

Another question the reader might ask is, why should maps between
branes and antibranes unravel into a linear complex as opposed to a more
general set of maps?  For example, why can one not have a configuration
that unravels to something of the form
\begin{displaymath}
\xymatrix{
{\cal E}_0 \ar[r] &
{\cal E}_1 \ar[r] &
{\cal E}_2 \ar[r] \ar@/_/[l] &
{\cal E}_3 \ar[r] &
{\cal E}_4 \ar[r] &
{\cal E}_5 \ar@/_3ex/[ll]
}
\end{displaymath}
?

In fact, such 
configurations are allowed physically, and also play an important
mathematical role.

Physically, if we add a boundary operator ${\cal O}$ of $U(1)_R$ charge
$n$, then $[G, {\cal O}]$ has charge $n-1$, so the boundaries it lies
between must have relative $U(1)_R$ charge $1-n$, and so give rise
to the `wrong-way' maps displayed above.

Adding such operators deforms the BRST operator
\begin{displaymath}
Q_{BRST} \: = \: \overline{\partial} \: + \:
\sum_i \phi_i^{\alpha \beta}
\end{displaymath}
and demanding that $Q_{BRST}^2 = 0$ now merely implies
\begin{displaymath}
\sum_i \overline{\partial} \phi_i \: + \:
\sum_{i,j} \phi_i \circ \phi_j \: = \: 0
\end{displaymath}

Complexes with wrong-way arrows of the form above, such that the maps
obey the condition stated above, are examples of ``generalized complexes''
used in \cite{bk} to define a technical improvement of ordinary
derived categories.  The relevance of \cite{bk} to physics was first
described in \cite{calin1,calin2,diac}.

\subsection{Cardy condition and Hirzebruch-Riemann-Roch}

Other aspects of the physics of the open string B model have also
been shown to have a mathematical understanding.
For example, the Cardy condition, which says that interpreting the
annulus diagram in terms of either closed or open string propagation,
has been shown by A.~Caldararu to be the same mathematically as the
Hirzebruch-Riemann-Roch index theorem:
\begin{displaymath}
\int_M \mbox{ch}({\cal E})^* \wedge \mbox{ch}({\cal F}) \wedge
\mbox{td}(TM) \: = \:
\sum_i (-)^i \mbox{dim }\mbox{Ext}^i_M({\cal E}, {\cal F})
\end{displaymath}

\subsection{Open problems}

Lest the reader get the impression that the connection between
derived categories and physics is well-understood, there are still
some very basic open problems that have never been solved.
For example:
\begin{enumerate}
\item One of the most basic problems is that we have glossed over
technical issues in dealing with bundles of rank greater than one.
For such bundles, it is not completely understood how their
curvature modifies the boundary conditions on the open strings.
Such boundary conditions will modify the arguments we have given for
tachyon vevs, and although we are optimistic that the modification
will not make essential changes to the argument, no one knows
for certain.
\item Anomaly cancellation in the open string B model implies that one
can only have open strings between some D-branes and not others.
At the moment, we can only describe the condition in special circumstances,
as the (unknown) boundary conditions above play a crucial role.
On the one hand, this condition seems to violate the spirit of the
arguments we have presented so far, and we would hope that some
additional physical effect (anomaly inflow, perhaps) modifies the
conclusion.  On the other hand, this anomaly cancellation condition
plays a role in understanding how Ext groups arise.
\item We understand how to associate D-branes to some sheaves,
but not to other sheaves.  A more comprehensive dictionary would be useful.
\end{enumerate}
Although we now have most of the puzzle pieces, so to speak,
a complete comprehensive physical understanding still does not exist.

\section{Stacks in physics}

\subsection{Introduction}

So far we have discussed the physical understanding of derived categories
as one application of the renormalization group:  
the mathematical process of localization on quasi-isomorphisms is
realized physically by worldsheet boundary renormalization group flow.

Another application is to the physical understanding of stacks\footnote{
In this talk by ``stack'' we mean Deligne-Mumford stacks and their
smooth analogues.},
where the renormalization group will play an analogous role in
washing out potential presentation-dependence.

Although in the mathematics literature the words ``stack'' and
``orbifold'' are sometimes used interchangeably, 
in the physics community the term ``orbifold'' has a much more
restrictive meaning:  global quotients by finite effectively-acting groups.
Most stacks can {\it not} be understood as global quotients by
finite effectively-acting groups.

Understanding physical properties of string orbifolds has led
physicists through many ideas regarding orbifolds (in the sense
used by physicists, global quotients by finite effectively-acting
groups).  For example, 
for a time, many thought that the properties of orbifolds were
somehow intrinsic to string theory or CFT.
Later, the fact that string orbifolds are well-behaved CFT's unlike
sigma models on quotient spaces was attributed to a notion of
``B fields at quotient singularities'' \cite{edstrings95},
 a notion that only made
sense in special cases.  Later still, because of the properties of
D-branes in orbifolds \cite{dgm}, many claimed that string orbifolds were
the same as strings propagating on resolutions of quotient spaces,
a notion that can not make sense for terminal singularities such as
${\bf C}^4/{\bf Z}_2$ or in some nonsupersymmetric orbifolds such as
$[{\bf C}/{\bf Z}_k]$.

It was proposed in \cite{estx1} that a better way of
understanding the physical properties of string orbifolds lie
just in thinking of terms of the geometry of stacks.
For readers who are not well acquainted with the notion,
the quotient stack $[X/G]$ encodes properties of orbifolds.
For example,
a function on the stack $[X/G]$ is a $G$-invariant function on $X$,
a metric on the stack $[X/G]$ is a $G$-invariant metric on $X$,
a bundle on the stack $[X/G]$ is a $G$-equivariant bundle on $X$,
and so forth.
Roughly, the stack $[X/G]$ is the same as the space $X/G$ except
over the singularities of $X/G$, where the stack has additional structure
that results in it being smooth while $X/G$ is singular.
For $G$ finite, a map $Y \rightarrow [X/G]$ is a pair consisting of
a principal $G$ bundle $P$ over $Y$ together with a $G$-equivariant
map $P \rightarrow X$.  The reader should recognize this as a
twisted sector (defined by $P$) together with a twisted map in that
sector.
In other words, summing over maps into the stack $[X/G]$, for $G$ finite,
is the same as summing over the data in the path integral description
of a string orbifolds.

Part of the proposal of \cite{estx1} was that the well-behavedness of
string orbifold CFT's, as opposed to sigma models on quotient spaces,
could be understood geometrically as stemming from the smoothness of
the corresponding stack.  In particular, the stack
$[X/G]$ is smooth even when the space $X/G$ is singular because of
fixed-points of $G$.
Another part of the proposal of \cite{estx1}
was a conjecture for how the ``B fields at quotient singularities''
could be understood mathematically in terms of stacks, though that
particular conjecture has since been contradicted.

However, at the time of \cite{estx1}, there were many unanswered questions.
First and foremost was a mismatch of moduli.
One of the first tests of any proposed mathematical model of part of physics
is whether the physical moduli match mathematical moduli -- this was
one of the reasons described for why sheaves are believed to be a good
model of D-branes, for example.  Unfortunately,
physical moduli of string orbifolds are not the same as mathematical
moduli of stacks.  For example, the stack $[{\bf C}^2/{\bf Z}_2]$
is rigid, it has no moduli, whereas physically the ${\bf Z}_2$
string orbifold of ${\bf C}^2$ does have moduli, which are understood
as deformations and resolutions of the quotient space ${\bf C}^2/{\bf Z}_2$.
If string orbifolds can be understood in terms of stacks,
then it will be the first known case in which mathematical moduli do not
match physical moduli.

Understanding this moduli mismatch was the source of much work over the
next few years.

Another problem at the time of \cite{estx1}
was the issue of understanding how to describe strings
on more general stacks.  Most stacks can not be understood as,
global quotients by finite effectively-acting groups, and so cannot be
understood in terms of orbifolds in the sense used by physicists.
One of the motivations for thinking about stacks was to introduce
a potentially new class of string compactifications, so understanding
strings on more general stacks was of interest.
At a more basic level, if the notion of strings on more general stacks
did not make sense, then that would call into question whether 
string orbifolds can be understood meaningfully in terms of (special) stacks.

\subsection{Strings on more general stacks}

How to understand strings on more general stacks?
Although most stacks can not be understood as global orbifolds by
finite groups, locally in patches they look like quotients by
finite not-necessarily-effectively-acting groups.
Unfortunately, that does not help us physically: only\footnote{
Very recently some progress has been made understanding CFT's 
perturbatively in terms of local
data on the target space \cite{edcdr},
but the work described there is only perturbative, not nonperturbative,
and does not suffice to define a CFT.} global
quotients are known to define CFT's, and only effectively-acting quotients
are well-understood.

So, we are back to the question of describing strings on more general
stacks.  It is a mathematical result that most\footnote{Experts are referred
to an earlier footnote where we define our usage of `stack.'
For the exceptions to the rule described above, 
{\it i.e.} for those few stacks not presentable in the form $[X/G]$
for some not necessarily finite, not necessarily effectively-acting $G$,
it is not currently known whether they define a CFT.
Even if it is possible to associate a CFT to them,
it is certainly not known how one would associate a CFT to them.}
stacks can be presented as a global quotient $[X/G]$
for some group, not necessarily finite and not necessarily effectively-acting.
To such a presentation, we can associate a $G$-gauged sigma model.

So, given a presentation of the correct form, we can get a physical theory.
Unfortunately, such presentations are not unique, and different
presentations lead to very different physics.
For example, following the dictionary above, 
$[{\bf C}^2/{\bf Z}_2]$ defines a conformally-invariant two-dimensional
theory.  However, that stack can also be presented as $[X/{\bf C}^{\times}]$
where $X = ( {\bf C}^2 \times {\bf C}^{\times} )/{\bf Z}_2$,
and to that presentation one associates a non-conformally invariant
two-dimensional theory, a $U(1)$-gauged sigma model.
As stacks, they are the same,
\begin{displaymath}
[ {\bf C}^2/{\bf Z}_2 ] \: = \: [X/{\bf C}^{\times}]
\end{displaymath}
but the corresponding physics is very different.

Thus, we have a potential presentation-dependence problem.
These problems are, again, analogous to those in understanding the
appearance of derived categories in physics.
There, to a given object in a derived category, one picks a representative
with a physical description (as branes/antibranes/tachyons),
just as here, given a stack, we must first pick a physically-realizable
presentation.
Every equivalence class of objects has at least one physically-realizable
representation; unfortunately, such representatives are not unique.
It is conjectured that different representatives give rise to the
same low-energy physics, via boundary renormalization group flow,
but only indirect tests are possible.

Here also, we conjecture that worldsheet renormalization group flow
takes different presentations of the same stack to the same CFT.
Unfortunately, just as in the case of derived categories,
worldsheet renormalization group flow cannot be followed explicitly,
and so we can not explicitly check such a claim.
Instead, we must rely on indirect tests.

At the least, one would like to check that the spectrum of massless
states is presentation-independent.  After all, in the case of
derived categories, this was one of the important checks we outlined that
the renormalization group respected localization on quasi-isomorphisms:
spectra computed in non-conformal presentations matched 
spectra computed in conformal presentations believed to be the endpoint
of renormalization group flow.

Unfortunately, no such test is possible here.  Massless spectra are
only explicitly computable for global quotients by finite groups,
and have only been well-understood in the past for global quotients by
finite effectively-acting groups.  For global quotients by nonfinite
groups, there has been a longstanding unsolved technical question of how to
compute massless spectra.  Since the theory is not conformal,
the spectrum cannot be computed in the usual fashion by 
enumerating vertex operators, and although there exist conjectures
for how to find massive representations of some such states
in special cases (gauged {\it linear} sigma models),
not even in such special cases does anyone have conjectures for
massive representations of all states, much less a systematic computation
method.  The last (unsuccessful) attempt appearing in print was
in \cite{edlargen}, and there has been little progress since then.

Thus, there is no way to tell if massless spectra are the same across
presentations.  On the other hand, a presentation-independent ansatz
for massless spectra makes predictions for massless spectra in
situations where they are not explicitly computable.

As alluded to earlier, another of the first indirect tests of
the presentation-independence claim is whether deformations of stacks
match deformations of corresponding CFT's.  In every other known example
of geometry applied to physics, mathematical deformations match physical
deformations.  Unfortunately, stacks fail this test,
which one might worry might be a signal of presentation-dependence.
Maybe renormalization group flow does not respect stacky equivalences
of gauged sigma models; maybe some different mathematics is
relevant instead of stacks.

To justify that stacks are relevant physically, as opposed to some
other mathematics, one has to understand this deformation theory issue,
as well as conduct tests for presentation-dependence.
This was the subject of several papers \cite{ps0,ps1,ps2}.

In the rest of these notes, we shall focus on special kinds of stacks
known as gerbes (which are described physically by quotients by
noneffectively-acting groups).

\subsection{Strings on gerbes}

Strings on gerbes, {\it i.e.}, global quotients by noneffectively-acting
groups, have additional physical difficulties beyond those mentioned
in the last subsection.
For example, the naive massless spectrum calculation contains
multiple dimension zero operators, which manifestly violates
cluster decomposition, one of the foundational axioms of quantum field theory.

There is a single known loophole:  if the target space is disconnected,
in which case cluster decomposition is also violated, but in the
mildest possible fashion.
We believe that is more or less what is going on.

Consider $[X/H]$ where
\begin{displaymath}
1 \: \longrightarrow \: G \: \longrightarrow \: H \: \longrightarrow \: K
\: \longrightarrow \: 1
\end{displaymath}
$G$ acts trivially, $K$ acts effectively, and neither $H$ nor $K$
need be finite.

We claim \cite{ps3}
\begin{displaymath}
\mbox{CFT}([X/H]) \: = \: \mbox{CFT}([(X \times \hat{G}) / K])
\end{displaymath}
(together with some $B$ field), where $\hat{G}$ is the set of
irreducible representations of $G$.
We refer to this as our ``decomposition conjecture.''
The stack $[( X \times \hat{G})/K]$ is not connected,
and so the CFT violates cluster decomposition but in the mildest
possible fashion.

For banded gerbes, $K$ acts trivially upon $\hat{G}$, so the
decomposition conjecture reduces to 
\begin{displaymath}
\mbox{CFT}\left( G-\mbox{gerbe on }X\right) \: = \:
\mbox{CFT}\left( \coprod_{\hat{G}} (X,B) \right)
\end{displaymath}
where the $B$ field on each component is determined by the image of
the characteristic class of the gerbe under
\begin{displaymath}
H^2(X, Z(G)) \: \stackrel{ Z(G) \rightarrow U(1) }{\longrightarrow} \:
H^2(X, U(1))
\end{displaymath}

For our first example, consider $[X/D_4]$, where the ${\bf Z}_2$ center 
of the dihedral group $D_4$ 
acts trivially:
\begin{displaymath}
1 \: \longrightarrow \: {\bf Z}_2 \: \longrightarrow \: D_4 \:
\longrightarrow \: {\bf Z}_2 \times {\bf Z}_2 \: \longrightarrow \: 1
\end{displaymath}
This example is banded, and the decomposition conjecture above predicts
\begin{displaymath}
\mbox{CFT}([X/D_4]) \: = \: \mbox{CFT}\left( [X/{\bf Z}_2 \times {\bf Z}_2]
\coprod [X/{\bf Z}_2 \times {\bf Z}_2] \right)
\end{displaymath}
One of the effective ${\bf Z}_2 \times {\bf Z}_2$ orbifolds has vanishing
discrete torsion, the other has nonvanishing discrete torsion,
using the relationship between discrete torsion and $B$ fields
first described in \cite{dt3,dtrev}.

One easy check of that statement lies in computing genus one partition
functions.
Denote the elements of the group $D_4$ by
\begin{displaymath}
D_4 \: = \: \{ 1, z, a, b, az, bz, ab, ba=abz \}
\end{displaymath}
and the elements of ${\bf Z}_2 \times {\bf Z}_2$ by
\begin{displaymath}
{\bf Z}_2 \times {\bf Z}_2 \: = \: \{ 1, \overline{a}, \overline{b},
\overline{ab} \}
\end{displaymath}
and the map $D_4 \rightarrow {\bf Z}_2 \times {\bf Z}_2$ proceeds as,
for example, $a, az \mapsto \overline{a}$.
The genus one partition function of the noneffective $D_4$ orbifold can
be described as
\begin{displaymath}
Z(D_4) \: = \: \frac{1}{| D_4 |} \sum_{g,h \in D_4, gh=hg}
Z_{g,h}
\end{displaymath}
Each of the $Z_{g,h}$ twisted sectors that appears is the same
as a ${\bf Z}_2 \times {\bf Z}_2$ sector, appearing with multiplicity
$| {\bf Z}_2 |^2 = 4$ except for the
\begin{displaymath}
{\scriptstyle \overline{a}} \square_{ \overline{b} }, \: \:
{\scriptstyle \overline{a}} \square_{ \overline{ab} }, \: \:
{\scriptstyle \overline{b}} \square_{ \overline{ab} }
\end{displaymath}
sectors, which have no lifts to the $D_4$ orbifold.
The partition function can be expressed as
\begin{eqnarray*}
Z(D_4) & = & \frac{ | {\bf Z}_2 \times {\bf Z}_2 | }{| D_4 |}
| {\bf Z}_2 |^2 \left( Z( {\bf Z}_2 \times {\bf Z}_2) \: - \:
\left( \mbox{some twisted sectors} \right) \right) \\
& = & 2 \left( Z({\bf Z}_2 \times {\bf Z}_2) \: - \:
\left(\mbox{some twisted sectors} \right) \right)
\end{eqnarray*}
The factor of 2 is important -- in ordinary QFT, one ignores multiplicative
factors in partition functions, but string theory is a two-dimensional
QFT coupled to gravity, and so such numerical factors are important.

Discrete torsion acts as a sign on the
\begin{displaymath}
{\scriptstyle \overline{a}} \square_{ \overline{b} }, \: \:
{\scriptstyle \overline{a}} \square_{ \overline{ab} }, \: \:
{\scriptstyle \overline{b}} \square_{ \overline{ab} }
\end{displaymath}
${\bf Z}_2 \times {\bf Z}_2$ twisted sectors, so adding partition functions
with and without discrete torsion will have the effect of removing the
sectors above and multiplying the rest by a factor of two.
Thus, we see that
\begin{displaymath}
Z([X/D_4]) \: = \: Z\left( [X/{\bf Z}_2 \times {\bf Z}_2] \coprod
[X/{\bf Z}_2 \times {\bf Z}_2] \right)
\end{displaymath}
with discrete torsion in one component.
(The same computation is performed at arbitrary genus in \cite{ps3}.)

Another quick tests of this example comes from comparing massless spectra.
Using the Hodge decomposition, the massless spectrum for $[T^6/D_4]$
can be expressed as
\begin{displaymath}
\begin{array}{ccccccc}
 & & & 2 & & & \\
 & & 0 & & 0 & & \\
 & 0 & & 54 & & 0 & \\
2 & & 54 & & 54 & & 2 \\
 & 0 & & 54 & & 0 & \\
 & & 0 & & 0 & & \\
 & & & 2 & & & \end{array}
\end{displaymath}
and the massless spectrum for each $[T^6/{\bf Z}_2 \times {\bf Z}_2]$,
with and without discrete torsion, can be written
\begin{displaymath}
\begin{array}{ccccccc}
 & & & 1 & & & \\
 & & 0 & & 0 & & \\
 & 0 & & 3 & & 0 & \\
1 & & 51 & & 51 & & 1 \\
 & 0 & & 3 & & 0 & \\
 & & 0 & & 0 & & \\
 & & & 1 & & & \end{array}
\end{displaymath}
and
\begin{displaymath}
\begin{array}{ccccccc}
 & & & 1 & & & \\
 & & 0 & & 0 & & \\
 & 0 & & 51 & & 0 & \\
1 & & 3 & & 3 & & 1 \\
 & 0 & & 51 & & 0 & \\
 & & 0 & & 0 & & \\
 & & & 1 & & & \end{array}
\end{displaymath}
The sum of the states from the two $[T^6/{\bf Z}_2 \times {\bf Z}_2]$ factors
matches that of $[T^6/D_4]$, precisely as expected.

Another example of the decomposition conjecture is given by
$[X/{\bf H}]$, where ${\bf H}$ is the eight-element group of 
quaternions, and a ${\bf Z}_4$ acts trivially:
\begin{displaymath}
1 \: \longrightarrow \: <i> (\cong {\bf Z}_4) \: \longrightarrow \:
{\bf H} \: \longrightarrow \: {\bf Z}_2 \: \longrightarrow \: 1
\end{displaymath}
The decomposition conjecture predicts
\begin{displaymath}
\mbox{CFT}([X/{\bf H}]) \: = \: \mbox{CFT}\left(
[X/{\bf Z}_2] \coprod [X/{\bf Z}_2] \coprod X \right)
\end{displaymath}
It is straightforward to show that this statement is true at the level
of partition functions, as before.

Another class of examples involves global quotients by noneffectively-acting
nonfinite groups.  For example, the banded ${\bf Z}_k$ gerbe over
${\bf P}^{N-1}$ with characteristic class $-1 \mbox{ mod }k$ can be
described mathematically as the quotient
\begin{displaymath}
\left[ \frac{ {\bf C}^{N} \: - \: \{0\} }{{\bf C}^{\times} } \right]
\end{displaymath}
where the ${\bf C}^{\times}$ acts as rotations by $k$ times rather than once.
Physically this quotient can be described by a $U(1)$ supersymmetric
gauge theory with $N$ chiral fields all of charge $k$, rather than
charge $1$.  The only difference between this and the ordinary
supersymmetric ${\bf P}^{N-1}$ model is that the charges are nonminimal.

Now, how can this be physically distinct from the ordinary supersymmetric
${\bf P}^{N-1}$ model?  After all, perturbatively having nonminimal
charges makes no difference.
The difference lies in nonperturbative effects.
For example, consider the anomalous global $U(1)$ symmetries of these
models.  In the ordinary supersymmetric ${\bf P}^{N-1}$ model,
the axial $U(1)$ is broken to ${\bf Z}_{2N}$ by anomalies, whereas
here it is broken to ${\bf Z}_{2kN}$.  The nonvanishing A model
correlation functions of the ordinary supersymmetric ${\bf P}^{N-1}$
model are given by
\begin{displaymath}
< X^{N(d+1)-1}> \: = \: q^d
\end{displaymath}
whereas here the nonzero A model correlation functions are given by
\begin{displaymath}
< X^{N(kd+1)-1}> \: = \: q^d
\end{displaymath}
As a result, the quantum cohomology ring of the ordinary
${\bf P}^{N-1}$ model is given by
\begin{displaymath}
{\bf C}[x]/(x^N \: - \: q)
\end{displaymath}
whereas the quantum cohomology ring of the current model is given by
\begin{displaymath}
{\bf C}[x]/(x^{kN} \: - \: q)
\end{displaymath}
In short, having nonminimal charges does lead to different physics.

Why should having nonminimal charges make a difference nonperturbatively?
On a compact worldsheet, this can be understood as follows.
To specify a Higgs field completely on a compact space, we need
to specify what bundle they couple to.  Thus, if the gauge field
couples to ${\cal L}$ then saying a Higgs field $\Phi$ has charge $Q$
implies $\Phi \in \Gamma({\cal L}^{\otimes Q})$.
Different bundles implies fields have different zero modes,
which implies different anomalies, which implies different physics.

On a noncompact worldsheet, the argument is different \cite{distpless}.
If electrons have charge $k$, then instantons have charge $1/k$,
and the theory reduces to the minimal-charge case.
Suppose we add massive fields of charge $\pm 1$, of mass greater than
the energy scale at which we are working.  One can 
determine instanton numbers by periodicity of the theta angle, which
acts like an electric field in two dimensions.  If all fields have charge $k$,
then the theta angle has periodicity $2 \pi k$, and we reduce to the ordinary
case.  However, the existence of massive fields of unit charge means
the theta angle has periodicity $2 \pi$, which is the new case.
Thus, even on a noncompact worldsheet, having nonminimal charges can
be distinguished from minimal charges.

There are four-dimensional analogues of this distinction.
For example, $SU(n)$ and $SU(n)/{\bf Z}_n$ gauge theories are 
perturbatively equivalent (since their Lie algebras are identical),
but have distinct nonperturbative corrections, a fact that is
crucial to the analysis of \cite{kaped}.
Similarly, $\mbox{Spin}(n)$ and $SO(n)$ gauge theories are
perturbatively identical but nonperturbatively distinct.
M. Strassler has studied Seiberg duality in this context
\cite{strassler1}, and has examples of $\mbox{Spin}(n)$ gauge
theories with ${\bf Z}_2$ monopoles (distinguishing 
$\mbox{Spin}(n)$ from $SO(n)$ nonperturbatively) Seiberg dual to
$\mbox{Spin}(n)$ gauge theory with massive spinors (distinguishing
$\mbox{Spin}(n)$ from $SO(n)$ perturbatively).

The equivalence of CFT's implied by our decomposition conjecture
implies a statement about K theory, thanks to D-branes.
Suppose $H$ acts on $X$ with a trivially-acting subgroup $G$:
\begin{displaymath}
1 \: \longrightarrow \:  G \: \longrightarrow \: H \: \longrightarrow \: K
\: \longrightarrow \: 1
\end{displaymath}
Our decomposition conjecture predicts that the ordinary $H$-equivariant
K theory of $X$ is the same as the twisted $K$-equivariant K theory
of $X \times \hat{G}$.  This result can be derived just within
K theory (see \cite{ps3}), and provides a check of the decomposition 
conjecture.

Another check of the decomposition conjecture comes from derived
categories.  Our decomposition conjecture predicts that D-branes
on a gerbe should be the same as D-branes on a disjoint union of
spaces, together with flat B fields, and this corresponds to a known
mathematics result.  Specifically, a sheaf on a gerbe is the
same as a twisted sheaf on $[X \times \hat{G}/K]$.
A sheaf on a banded $G$-gerbe is the same thing as a twisted
sheaf on the underlying space, twisted by the image of 
the characteristic class of the gerbe in $H^2(X, Z(G))$.
Thus, sheaves on gerbes behave in exactly the fashion one would
expect from D-branes according to our decomposition conjecture.

Similarly, massless states between D-branes also have an analogous
decomposition.  For D-branes on a disjoint union of spaces, there will
only be massless states between D-branes which are both on the same
connected component.  Mathematically, in the banded case for example,
sheaves on a banded $G$-gerbe decompose according to irreducible
representations of $G$, and sheaves associated to distinct irreducible
representations have vanishing Ext groups between them.
This is precisely consistent with the idea that sheaves associated to
distinct irreducible representations should describe D-branes on
different components of a disconnected space.

\subsection{Mirror symmetry for stacks}

There exist mirror constructions for any model realizable as
a two-dimensional abelian gauge theory \cite{mpmirror,hv}.
There is a notion of toric stacks \cite{bcs}, generalizing
toric varieties, which can be described physically
via gauged linear sigma models \cite{ps2}.  
Standard mirror constructions \cite{mpmirror,hv} now produce \cite{ps2}
character-valued fields, a new effect, which ties into
the stacky fan description of \cite{bcs}.

For example, the ``Toda dual'' of the supersymmetric
${\bf P}^{N}$ model is described by Landau-Ginzburg model
with superpotential
\begin{displaymath}
W \: = \: \exp(-Y_1) \: + \: \cdots \: + \: 
\exp(-Y_N) \: + \: \exp(Y_1 \: + \: \cdots \: + \: Y_N)
\end{displaymath}
The analogous duals to ${\bf Z}_k$ gerbes over ${\bf P}^N$,
of characteristic class $-n \bmod k$, are given by \cite{ps2}
\begin{displaymath}
W \: = \: \exp(-Y_1) \: + \: \cdots \: + \: \exp(-Y_N) \: + \:
\Upsilon^n \exp(Y_1 \: + \: \cdots \: + \: Y_N)
\end{displaymath}
where $\Upsilon$ is a character-valued field, in this case
valued in the characters of ${\bf Z}_k$.

In the same language, the Landau-Ginzburg point mirror to
the quintic hypersurface in a ${\bf Z}_k$ gerbe over ${\bf P}^4$
is described by (an orbifold of) the superpotential
\begin{displaymath}
W \: = \: x_0^5 \: + \: \cdots \: + \: x_4^5 \: + \:
\psi \Upsilon x_0 x_1 x_2 x_3 x_4
\end{displaymath}
where $\psi$ is the ordinary complex structure parameter
(mirror to the K\"ahler parameter), and $\Upsilon$ is a discrete
(character-)valued field as above.

In terms of the path integral measure,
\begin{displaymath}
\int[ D x_i, \Upsilon] \: = \: \int [D x_i] \sum_{\Upsilon} \: = \:
\sum_{\Upsilon} \int [D x_i]
\end{displaymath}
so having a discrete-valued field is equivalent to summing over contributions
from different theories, or, equivalently, summing over different
components of the target space.

In the case of the gerby quintic, the presence of the discrete-valued
field $\Upsilon$ on the mirror means that the CFT is describing
a target space with multiple components.
Moreover,
the mirror map for the ordinary quintic says
\begin{displaymath}
B \: + \: i J \: = \: - \frac{5}{2 \pi i} \log(5 \psi) \: + \: \cdots
\end{displaymath}
so shifting $\psi$ by phases has precisely\footnote{Higher order corrections
invalidate geometric conclusions, so we are omitting them.} 
the effect of shifting the B field, exactly as the decomposition conjecture
predicts for this case.

\subsection{Applications}

One of the original proposed applications of these ideas,
described in \cite{estx1},
was to understand physical properties of string orbifolds.
For example, the fact that string orbifolds define well-behaved
CFT's, unlike sigma models on quotient spaces, might be attributable
to the smoothness of the corresponding quotient stack,
instead of traditional notions such as ``B fields at quotient singularities''
or ``string orbifolds are strings on resolutions,''
which do not even make sense in general.

Another basic application is to give a concrete understanding
of local orbifolds, {\it i.e.}, stacks which are locally quotients
by finite groups but which cannot be expressed globally as quotients
by finite groups.  We now have a concrete way to manufacture a 
corresponding CFT -- by rewriting the local orbifold as a global
quotient by a nonfinite group -- and we also understand what problems
may arise -- the construction might not be well-defined, as different
rewritings might conceivably flow under the renormalization group to
distinct CFT's.

Implicit here is that stacks give a classification of universality classes
of worldsheet renormalization group flow in gauged sigma models,
just as derived categories give a classification of universality classes
of worldsheet boundary renormalization group flow in the open string
B model.

Another application is the computation of massless spectra in cases
where direct calculations are not currently possible,
such as global quotients by nonfinite groups.  A presentation-independent
ansatz was described in \cite{estx1,ps0,ps1,ps2} which predicts massless
spectra in cases where explicitly enumerating vertex operators is not
possible.

Another application of these ideas is to the properties of 
quotients by noneffective group actions, {\it i.e.} group actions in
which elements other than the identity act trivially.
Such quotients correspond to strings propagating on special kinds of
stacks known as gerbes.

Noneffective group actions play a crucial role in \cite{kaped},
the recent work on the physical interpretation of the geometric
Langlands program.  One application of the decomposition conjecture
of the last section is to give a concrete understanding of
some aspects of \cite{kaped}.  For example,
related work \cite{dpgl} describes two-dimensional theories
in the language of gerbes,
whereas \cite{kaped} deals exclusively with spaces.  
As a result of the decomposition conjecture, we see that 
the language of \cite{dpgl} is physically equivalent to that of 
\cite{kaped}, as sigma models on the gerbes of \cite{dpgl} define
the same CFT's as sigma models on the disjoint unions of spaces
of \cite{kaped}.

Our decomposition conjecture  makes a prediction for Gromov-Witten invariants of
gerbes, as defined in the mathematics literature in,
for example, \cite{agv}.  Specifically, the Gromov-Witten theory of
$[X/H]$ should match that of $[(X \times \hat{G})/K]$.
This prediction works in basic cases \cite{graberbryan}.

Another result of our work is quantum cohomology for toric stacks.
Toric stacks are a stacky generalization of toric varieties
\cite{bcs}.  Just as toric varieties can be described with
gauged linear sigma models, so too can toric stacks
\cite{ps2}, and so the technology of gauged linear sigma models
can be applied to their understanding.
In particular, Batyrev's conjecture for quantum cohomology rings
can be extracted from the two-dimensional effective action of the
gauge theory, without any explicit mention of rational curves
\cite{morrisonpless}.

In the present case, old results of \cite{morrisonpless} generalize
from toric varieties to toric stacks.
Let the toric stack be described in the form
\begin{displaymath}
\left[ \frac{ {\bf C}^N \: - \: E }{ ( {\bf C}^{\times} )^n } \right]
\end{displaymath}
where $E$ is some exceptional set, $Q_i^a$ the weight of the
$i$th vector under the $a$th ${\bf C}^{\times}$,
then the analogue of Batyrev's conjecture for the quantum cohomology ring
is of the form ${\bf C}[\sigma_1, \cdots, \sigma_n]$ modulo the
relations \cite{ps2}
\begin{displaymath}
\prod_{i=1}^N \left( \sum_{b=1}^n Q_i^b \sigma_b \right)^{Q^a_i}\: = \:
q_a
\end{displaymath}

For example, the quantum cohomology ring of ${\bf P}^N$ is
\begin{displaymath}
{\bf C}[x]/(x^{N+1} \: - \: q)
\end{displaymath}
and according to the formula above the quantum cohomology ring
of a ${\bf Z}_k$ gerbe over ${\bf P}^N$ with characteristic class
$-n \mbox{ mod } k$ is
\begin{displaymath}
{\bf C}[x,y]/(y^k \: - \: q_2, \: \: x^{N+1} \: - \: y^n q_1)
\end{displaymath}

As an aside, note the calculations above give us a check
of the massless spectrum.  In physics, we can derive quantum cohomology
rings without knowing the massless spectrum, and we are unable to
calculate the massless spectrum directly for the gerbes above,
hence we can use the quantum cohomology rings to read off the
additive part of the massless spectrum.

Also note that we can see the decomposition conjecture for gerbes
in the quantum cohomology rings of toric stacks.
Consider for example the quantum cohomology ring of a ${\bf Z}_k$
gerbe on ${\bf P}^N$, as above.  In that ring, the $y$'s index copies
of the quantum cohomology ring of ${\bf P}^N$ with variable $q$'s.
The gerbe is banded, so this is exactly what we expect -- copies of
${\bf P}^N$, variable B field.

More generally, a gerbe structure is indicated from the quotient description
whenever the ${\bf C}^{\times}$ charges are nonminimal.  In such a case,
from our generalization of Batyrev's conjecture, at least one relation
will have the form $p^k = q$, where $p$ is a relation in the quantun cohomology
ring of the toric variety, and $k$ is the greatest divisor in the
nonminimal charges.  We can rewrite that relation in the same form as for
a gerbe on ${\bf P}^N$, and in this fashion can see our decomposition 
conjecture in our generalization of Batyrev's quantum cohomology.

Other applications of stacks to understanding D-branes and their
derived categories model are discussed in \cite{karp1,karp2,karp3}.

\section{Conclusions}

In this talk we have outlined how both derived categories and stacks
enter physics, and the crucial role of the renormalization group.
In both cases, to physically realize either a derived category
or stack, one picks physically-realizable presentations
(which are guaranteed to exist, though not all presentations are
physically realizable), which yield nonconformal theories.
Remaining presentation-dependence is
removed via renormalization group flow.

\section{Acknowledgements}

I have learned a great deal of both mathematics and physics from
my collaborators on the papers we have written concerning
derived categories and
stacks.
Listed alphabetically, they are
M.~Ando, A.~Caldararu, R.~Donagi, S.~Hellerman, A.~Henriques, S.~Katz, 
and T.~Pantev.

\end{document}